\documentclass[runningheads]{llncs}
\usepackage[T1]{fontenc}
\usepackage{graphicx}
\usepackage{acronym}
\usepackage{amsmath}
\usepackage{amssymb}
\usepackage{amsfonts}
\usepackage[table,xcdraw]{xcolor}
\usepackage{xspace}
\usepackage{booktabs}
\usepackage{caption}
\usepackage{subcaption}
\usepackage{paralist}
\usepackage[inline]{enumitem}

\acrodef{3G}[3G]{Third Generation Mobile System}
\acrodef{5S}[5S]{Streams, Structures, Spaces, Scenarios, Societies}
\acrodef{AAAI}[AAAI]{Association for the Advancement of Artificial Intelligence}
\acrodef{AAL}[AAL]{Annotation Abstraction Layer}
\acrodef{AAM}[AAM]{Automatic Annotation Manager}
\acrodef{AAP}{Average Average Precision}
\acrodef{ACLIA}[ACLIA]{Advanced Cross-Lingual Information Access}
\acrodef{ACM}[ACM]{Association for Computing Machinery}
\acrodef{ADSL}[ADSL]{Asymmetric Digital Subscriber Line}
\acrodef{ADUI}[ADUI]{ADministrator User Interface}
\acrodef{AIP}[AIP]{Archival Information Package}
\acrodef{AJAX}[AJAX]{Asynchronous JavaScript Technology and \acs{XML}}
\acrodef{ALU}[ALU]{Aritmetic-Logic Unit}
\acrodef{AMUSID}[AMUSID]{Adaptive MUSeological IDentity-service}
\acrodef{ANOVA}[ANOVA]{ANalysis Of VAriance}
\acrodef{ANSI}[ANSI]{American National Standards Institute}
\acrodef{AP}[AP]{Average Precision}
\acrodef{APC}[APC]{AP Correlation}
\acrodef{API}[API]{Application Program Interface}
\acrodef{AR}[AR]{Address Register}
\acrodef{AS}[AS]{Annotation Service}
\acrodef{ASAP}[ASAP]{Adaptable Software Architecture Performance}
\acrodef{ASI}[ASI]{Annotation Service Integrator}
\acrodef{ASL}[ASL]{Achieved Significance Level}
\acrodef{ASM}[ASM]{Annotation Storing Manager}
\acrodef{ASR}[ASR]{Automatic Speech Recognition}
\acrodef{ASUI}[ASUI]{ASsessor User Interface}
\acrodef{ATIM}[ATIM]{Annotation Textual Indexing Manager}
\acrodef{AUC}[AUC]{Area Under the ROC Curve}
\acrodef{AUI}[AUI]{Administrative User Interface}
\acrodef{AWARE}[AWARE]{Assessor-driven Weighted Averages for Retrieval Evaluation}
\acrodef{BANKS-I}[BANKS-I]{Browsing ANd Keyword Searching I}
\acrodef{BAKS-II}[BANKS-II]{Browsing ANd Keyword Searching II}
\acrodef{BH}[BH]{Benjamini Hochberg}
\acrodef{bpref}[bpref]{Binary Preference}
\acrodef{BNF}[BNF]{Backus and Naur Form}
\acrodef{BRICKS}[BRICKS]{Building Resources for Integrated Cultural Knowledge Services}
\acrodef{CAN}[CAN]{Content Addressable Netword}
\acrodef{CAS}[CAS]{Content-And-Structure}
\acrodef{CBSD}[CBSD]{Component-Based Software Developlement}
\acrodef{CBSE}[CBSE]{Component-Based Software Engineering}
\acrodef{CB-SPE}[CB-SPE]{Component-Based \acs{SPE}}
\acrodef{CD}[CD]{Collaboration Diagram}
\acrodef{CD}[CD]{Compact Disk}
\acrodef{CF}[CF]{Collection Frequency}
\acrodef{CDF}[CDF]{Cumulative Density Function}
\acrodef{CENL}[CENL]{Conference of European National Librarians}
\acrodef{CIDOC CRM}[CIDOC CRM]{CIDOC Conceptual Reference Model}
\acrodef{CIR}[CIR]{Current Instruction Register}
\acrodef{CIRCO}[CIRCO]{Coordinated Information Retrieval Components Orchestration}
\acrodef{CG}[CG]{Cumulated Gain}
\acrodef{CI}[CI]{Confidence Interval}
\acrodef{CL}[CL]{Curriculum Learning}
\acrodef{CL-ESA}[CL-ESA]{Cross-Lingual Explicit Semantic Analysis}
\acrodef{CLAIRE}[CLAIRE]{Combinatorial visuaL Analytics system for Information Retrieval Evaluation}
\acrodef{CLEF1}[CLEF]{Cross-Language Evaluation Forum}
\acrodef{CLEF}[CLEF]{Conference and Labs of the Evaluation Forum}
\acrodef{CLIR}[CLIR]{Cross Language Information Retrieval}
\acrodef{CM}[CM]{Continuation Methods}
\acrodef{CMS}[CMS]{Content Management System}
\acrodef{CMT}[CMT]{Campaign Management Tool}
\acrodef{CNR}[CNR]{Italian National Council of Research}
\acrodef{CO}[CO]{Content-Only}
\acrodef{COD}[COD]{Code On Demand}
\acrodef{CODATA}[CODATA]{Committee on Data for Science and Technology}
\acrodef{ColBERT}[ColBERT]{Contextualized late interaction over BERT}
\acrodef{COLLATE}[COLLATE]{Collaboratory for Annotation Indexing and Retrieval of Digitized Historical Archive Material}
\acrodef{CP}[CP]{Characteristic Pattern}
\acrodef{CPE}[CPE]{Control Processor Element}
\acrodef{CPU}[CPU]{Central Processing Unit}
\acrodef{CQL}[CQL]{Contextual Query Language}
\acrodef{CRP}[CRP]{Cumulated Relative Position}
\acrodef{CRUD}[CRUD]{Create--Read--Update--Delete}
\acrodef{CS}[CS]{Characteristic Structure}
\acrodef{CSM}[CSM]{Campaign Storing Manager}
\acrodef{CSS}[CSS]{Cascading Style Sheets}
\acrodef{CTR}[CTR]{Click-Through Rate}
\acrodef{CU}[CU]{Control Unit}
\acrodef{CUI}[CUI]{Client User Interface}
\acrodef{CV}[CV]{Cross-Validation}
\acrodef{DAFFODIL}[DAFFODIL]{Distributed Agents for User-Friendly Access of Digital Libraries}
\acrodef{DAO}[DAO]{Data Access Object}
\acrodef{DARE}[DARE]{Drawing Adequate REpresentations}
\acrodef{DARPA}[DARPA]{Defense Advanced Research Projects Agency}
\acrodef{DAS}[DAS]{Distributed Annotation System}
\acrodef{DB}[DB]{DataBase}
\acrodef{DBMS}[DBMS]{DataBase Management System}
\acrodef{DC}[DC]{Dublin Core}
\acrodef{DCG}[DCG]{Discounted Cumulated Gain}
\acrodef{DCMI}[DCMI]{Dublin Core Metadata Initiative}
\acrodef{DCV}[DCV]{Document Cut--off Value}
\acrodef{DD}[DD]{Deployment Diagram}
\acrodef{DDC}[DDC]{Dewey Decimal Classification}
\acrodef{DDS}[DDS]{Direct Data Structure}
\acrodef{DF}[DF]{Degrees of Freedom}
\acrodef{DFI}[DFI]{Divergence From Independence}
\acrodef{DFR}[DFR]{Divergence From Randomness}
\acrodef{DHT}[DHT]{Distributed Hash Table}
\acrodef{DI}[DI]{Digital Image}
\acrodef{DIKW}[DIKW]{Data, Information, Knowledge, Wisdom}
\acrodef{DIL}[DIL]{\acs{DIRECT} Integration Layer}
\acrodef{DiLAS}[DiLAS]{Digital Library Annotation Service}
\acrodef{DIRECT}[DIRECT]{Distributed Information Retrieval Evaluation Campaign Tool}
\acrodef{DKMS}[DKMS]{Data and Knowledge Management System}
\acrodef{DL}[DL]{Digital Library}
\acrodefplural{DL}[DL]{Digital Libraries}
\acrodef{DLMS}[DLMS]{Digital Library Management System}
\acrodef{DLOG}[DL]{Description Logics}
\acrodef{DLS}[DLS]{Digital Library System}
\acrodef{DLSS}[DLSS]{Digital Library Service System}
\acrodef{DM}[DM]{Data Mining}
\acrodef{DO}[DO]{Digital Object}
\acrodef{DOI}[DOI]{Digital Object Identifier}
\acrodef{DOM}[DOM]{Document Object Model}
\acrodef{DoMDL}[DoMDL]{Document Model for Digital Libraries}
\acrodef{DP}[DP]{Discriminative Power}
\acrodef{DPBF}[DPBF]{Dynamic Programming Best-First}
\acrodef{DR}[DR]{Data Register}
\acrodef{DRIVER}[DRIVER]{Digital Repository Infrastructure Vision for European Research}
\acrodef{DRMM}[DRMM]{Deep Relevance Matching Model}
\acrodef{DTD}[DTD]{Document Type Definition}
\acrodef{DVD}[DVD]{Digital Versatile Disk}
\acrodef{EAC-CPF}[EAC-CPF]{Encoded Archival Context for Corporate Bodies, Persons, and Families}
\acrodef{EAD}[EAD]{Encoded Archival Description}
\acrodef{EAN}[EAN]{International Article Number}
\acrodef{EBU}[EBU]{Expected Browsing Utility}
\acrodef{ECD}[ECD]{Enhanced Contenty Delivery}
\acrodef{ECDL}[ECDL]{European Conference on Research and Advanced Technology for Digital Libraries}
\acrodef{EDM}[EDM]{Europeana Data Model}
\acrodef{EG}[EG]{Execution Graph}
\acrodef{ELDA}[ELDA]{Evaluation and Language resources Distribution Agency}
\acrodef{ELRA}[ELRA]{European Language Resources Association}
\acrodef{EM}[EM]{Expectation Maximization}
\acrodef{EMMA}[EMMA]{Extensible MultiModal Annotation}
\acrodef{EPROM}[EPROM]{Erasable Programmable \acs{ROM}}
\acrodef{EQNM}[EQNM]{Extended Queueing Network Model}
\acrodef{ERR}[ERR]{Expected Reciprocal Rank}
\acrodef{ESA}[ESA]{Explicit Semantic Analysis}
\acrodef{ESL}[ESL]{Expected Search Length}
\acrodef{ETL}[ETL]{Extract-Transform-Load}
\acrodef{FAST}[FAST]{Flexible Annotation Service Tool}
\acrodef{FDR}[FDR]{False Discovery Rate}
\acrodef{FIFO}[FIFO]{First-In / First-Out}
\acrodef{FIRE}[FIRE]{Forum for Information Retrieval Evaluation}
\acrodef{FN}[FN]{False Negative}
\acrodef{FNR}[FNR]{False Negative Rate}
\acrodef{FOAF}[FOAF]{Friend of a Friend}
\acrodef{FORESEE}[FORESEE]{FOod REcommentation sErvER}
\acrodef{FP}[FP]{False Positive}
\acrodef{FPR}[FPR]{False Positive Rate}
\acrodef{FWER}[FWER]{Family-wise Error Rate}
\acrodef{GIF}[GIF]{Graphics Interchange Format}
\acrodef{GIR}[GIR]{Geografic Information Retrieval}
\acrodef{GAP}[GAP]{Graded Average Precision}
\acrodef{GLM}[GLM]{Generalized Linear Model}
\acrodef{GLMM}[GLMM]{General Linear Mixed Model}
\acrodef{GLiM}[GLiM]{General Linear Model}
\acrodef{GMAP}[GMAP]{Geometric Mean Average Precision}
\acrodef{GMM}[GMM]{Gaussian Mixture Model}
\acrodef{GoP}[GoP]{Grid of Points}
\acrodef{GPRS}[GPRS]{General Packet Radio Service}
\acrodef{gP}[gP]{Generalized Precision}
\acrodef{gR}[gR]{Generalized Recall}
\acrodef{gRBP}[gRBP]{Graded Rank-Biased Precision}
\acrodef{GT}[GT]{Generalizability Theory}
\acrodef{GTIN}[GTIN]{Global Trade Item Number}
\acrodef{GUI}[GUI]{Graphical User Interface}
\acrodef{GW}[GW]{Gateway}
\acrodef{HCI}[HCI]{Human Computer Interaction}
\acrodef{HDS}[HDS]{Hybrid Data Structure}
\acrodef{HIR}[HIR]{Hypertext Information Retrieval}
\acrodef{HIT}[HIT]{Human Intelligent Task}
\acrodef{HITS}[HITS]{Hyperlink-Induced Topic Search}
\acrodef{HMM}[HMM]{Hidden Markov Model}
\acrodef{HTML}[HTML]{HyperText Markup Language}
\acrodef{HTTP}[HTTP]{HyperText Transfer Protocol}
\acrodef{HSD}[HSD]{Honestly Significant Difference}
\acrodef{ICA}[ICA]{International Council on Archives}
\acrodef{ICTF}[ICTF]{Inverse Collection Term Frequency}
\acrodef{ICSU}[ICSU]{International Council for Science}
\acrodef{IDF}[IDF]{Inverse Document Frequency}
\acrodef{IDS}[IDS]{Inverse Data Structure}
\acrodef{IEEE}[IEEE]{Institute of Electrical and Electronics Engineers}
\acrodef{IEI}[IEI]{Istituto della Enciclopedia Italiana fondata da Giovanni Treccani}
\acrodef{IETF}[IETF]{Internet Engineering Task Force}
\acrodef{IIR}[IIR]{Interactive Information Retrieval}
\acrodef{IMS}[IMS]{Information Management System}
\acrodef{IMSPD}[IMS]{Information Management Systems Research Group}
\acrodef{indAP}[indAP]{Induced Average Precision}
\acrodef{infAP}[infAP]{Inferred Average Precision}
\acrodef{INEX}[INEX]{INitiative for the Evaluation of \acs{XML} Retrieval}
\acrodef{INS-M}[INS-M]{Inverse Set Data Model}
\acrodef{INTR}[INTR]{Interrupt Register}
\acrodef{IP}[IP]{Internet Protocol}
\acrodef{IPSA}[IPSA]{Imaginum Patavinae Scientiae Archivum}
\acrodef{IR}[IR]{Information Retrieval}
\acrodef{IRON}[IRON]{Information Retrieval ON}
\acrodef{IRON2}[IRON$^2$]{Information Retrieval On aNNotations}
\acrodef{IRON-SAT}[IRON-SAT]{\acs{IRON} - Statistical Analysis Tool}
\acrodef{IRS}[IRS]{Information Retrieval System}
\acrodef{ISAD(G)}[ISAD(G)]{International Standard for Archival Description (General)}
\acrodef{ISBN}[ISBN]{International Standard Book Number}
\acrodef{ISIS}[ISIS]{Interactive SImilarity Search}
\acrodef{ISJ}[ISJ]{Interactive Searching and Judging}
\acrodef{ISO}[ISO]{International Organization for Standardization}
\acrodef{ITU}[ITU]{International Telecommunication Union }
\acrodef{ITU-T}[ITU-T]{Telecommunication Standardization Sector of \acs{ITU}}
\acrodef{IV}[IV]{Information Visualization}
\acrodef{JAN}[JAN]{Japanese Article Number}
\acrodef{JDBC}[JDBC]{Java DataBase Connectivity}
\acrodef{JMB}[JMB]{Java--Matlab Bridge}
\acrodef{JPEG}[JPEG]{Joint Photographic Experts Group}
\acrodef{JSON}[JSON]{JavaScript Object Notation}
\acrodef{JSP}[JSP]{Java Server Pages}
\acrodef{JTE}[JTE]{Java-Treceval Engine}
\acrodef{KDE}[KDE]{Kernel Density Estimation}
\acrodef{KLD}[KLD]{Kullback-Leibler Divergence}
\acrodef{KLAPER}[KLAPER]{Kernel LAnguage for PErformance and Reliability analysis}
\acrodef{KPI}[KPI]{Key Performance Indicator}
\acrodef{LAM}[LAM]{Libraries, Archives, and Museums}
\acrodef{LAM2}[LAM]{Logistic Average Misclassification}
\acrodef{LAN}[LAN]{Local Area Network}
\acrodef{LD}[LD]{Linked Data}
\acrodef{LEAF}[LEAF]{Linking and Exploring Authority Files}
\acrodef{LIDO}[LIDO]{Lightweight Information Describing Objects}
\acrodef{LIFO}[LIFO]{Last-In / First-Out}
\acrodef{LM}[LM]{Language Model}
\acrodef{LMT}[LMT]{Log Management Tool}
\acrodef{LOD}[LOD]{Linked Open Data}
\acrodef{LODE}[LODE]{Linking Open Descriptions of Events}
\acrodef{LpO}[LpO]{Leave-$p$-Out}
\acrodef{LRM}[LRM]{Local Relational Model}
\acrodef{LRU}[LRU]{Last Recently Used}
\acrodef{LS}[LS]{Lexical Signature}
\acrodef{LSM}[LSM]{Log Storing Manager}
\acrodef{LtR}[LtR]{Learning to Rank}
\acrodef{LUG}[LUG]{Lexical Unit Generator}
\acrodef{MA}[MA]{Mobile Agent}
\acrodef{MA}[MA]{Moving Average}
\acrodef{MACS}[MACS]{Multilingual ACcess to Subjects}
\acrodef{MADCOW}[MADCOW]{Multimedia Annotation of Digital Content Over the Web}
\acrodef{MAD}[MAD]{Mean Assessed Documents}
\acrodef{MADP}[MADP]{Mean Assessed Documents Precision}
\acrodef{MADS}[MADS]{Metadata Authority Description Standard}
\acrodef{MAP}[MAP]{Mean Average Precision}
\acrodef{MARC}[MARC]{Machine Readable Cataloging}
\acrodef{MATTERS}[MATTERS]{MATlab Toolkit for Evaluation of information Retrieval Systems}
\acrodef{MDA}[MDA]{Model Driven Architecture}
\acrodef{MDD}[MDD]{Model-Driven Development}
\acrodef{METS}[METS]{Metadata Encoding and Transmission Standard}
\acrodef{MIDI}[MIDI]{Musical Instrument Digital Interface}
\acrodef{MIME}[MIME]{Multipurpose Internet Mail Extensions}
\acrodef{ML}[ML]{Machine Learning}
\acrodef{MLE}[MLE]{Maximum Likelihood Estimation}
\acrodef{MLIA}[MLIA]{MultiLingual Information Access}
\acrodef{MM}[MM]{Machinery Model}
\acrodef{MMU}[MMU]{Memory Management Unit}
\acrodef{MODS}[MODS]{Metadata Object Description Schema}
\acrodef{MOF}[MOF]{Meta-Object Facility}
\acrodef{MP}[MP]{Markov Precision}
\acrodef{MPEG}[MPEG]{Motion Picture Experts Group}
\acrodef{MRD}[MRD]{Machine Readable Dictionary}
\acrodef{MRF}[MRF]{Markov Random Field}
\acrodef{MRR}[MRR]{Mean Reciprocal Rank}
\acrodef{MS}[MS]{Mean Squares}
\acrodef{MSAC}[MSAC]{Multilingual Subject Access to Catalogues}
\acrodef{MSE}[MSE]{Mean Square Error}
\acrodef{MT}[MT]{Machine Translation}
\acrodef{MV}[MV]{Majority Vote}
\acrodef{MVC}[MVC]{Model-View-Controller}
\acrodef{NACSIS}[NACSIS]{NAtional Center for Science Information Systems}
\acrodef{NAP}[NAP]{Network processors Applications Profile}
\acrodef{NCP}[NCP]{Normalized Cumulative Precision}
\acrodef{nCG}[nCG]{Normalized Cumulated Gain}
\acrodef{nCRP}[nCRP]{Normalized Cumulated Relative Position}
\acrodef{nDCG}[nDCG]{Normalized Discounted Cumulated Gain}
\acrodef{nMCG}[nMCG]{Normalized Markov Cumulated Gain}
\acrodef{NESTOR}[NESTOR]{NEsted SeTs for Object hieRarchies}
\acrodef{NEXI}[NEXI]{Narrowed Extended XPath I}
\acrodef{NII}[NII]{National Institute of Informatics}
\acrodef{NIR}[NIR]{Neural IR}
\acrodef{NISO}[NISO]{National Information Standards Organization}
\acrodef{NIST}[NIST]{National Institute of Standards and Technology}
\acrodef{NLP}[NLP]{Natural Language Processing}
\acrodef{NN}[NN]{Neural Network}
\acrodef{NP}[NP]{Network Processor}
\acrodef{NQC}[NQC]{Normalized Query Commitment}
\acrodef{NR}[NR]{Normalized Recall}
\acrodef{NS-M}[NS-M]{Nested Set Model}
\acrodef{NTCIR}[NTCIR]{NII Testbeds and Community for Information access Research}
\acrodef{OAI}[OAI]{Open Archives Initiative}
\acrodef{OAI-ORE}[OAI-ORE]{Open Archives Initiative Object Reuse and Exchange}
\acrodef{OAI-PMH}[OAI-PMH]{Open Archives Initiative Protocol for Metadata Harvesting}
\acrodef{OAIS}[OAIS]{Open Archival Information System}
\acrodef{OC}[OC]{Operation Code}
\acrodef{OCLC}[OCLC]{Online Computer Library Center}
\acrodef{OLS}[OLS]{Ordinary Least Squares}
\acrodef{OMG}[OMG]{Object Management Group}
\acrodef{OO}[OO]{Object Oriented}
\acrodef{OODB}[OODB]{Object-Oriented \acs{DB}}
\acrodef{OODBMS}[OODBMS]{Object-Oriented \acs{DBMS}}
\acrodef{OPAC}[OPAC]{Online Public Access Catalog}
\acrodef{OQL}[OQL]{Object Query Language}
\acrodef{ORP}[ORP]{Open Relevance Project}
\acrodef{OSIRIS}[OSIRIS]{Open Service Infrastructure for Reliable and Integrated process Support}
\acrodef{P}[P]{Precision}
\acrodef{P2P}[P2P]{Peer-To-Peer}
\acrodef{PA}[PA]{Performance Analysis}
\acrodef{PAMT}[PAMT]{Pool-Assessment Management Tool}
\acrodef{PASM}[PASM]{Pool-Assessment Storing Manager}
\acrodef{PC}[PC]{Program Counter}
\acrodef{PCP}[PCP]{Pre-Commercial Procurement}
\acrodef{PCR}[PCR]{Peripherical Command Register}
\acrodef{PDA}[PDA]{Personal Digital Assistant}
\acrodef{PDF}[PDF]{Probability Density Function}
\acrodef{PDR}[PDR]{Peripherical Data Register}
\acrodef{PIR}[PIR]{Personalized Information Retrieval}
\acrodef{PLM}[PLM]{Pre-trained Language Models}
\acrodef{POI}[POI]{\acs{PURL}-based Object Identifier}
\acrodef{PoS}[PoS]{Part of Speech}
\acrodef{PPE}[PPE]{Programmable Processing Engine}
\acrodef{PREFORMA}[PREFORMA]{PREservation FORMAts for culture information/e-archives}
\acrodef{PRIMAD}[PRIMAD]{Platform, Research goal, Implementation, Method, Actor, and Data}
\acrodef{PRIMAmob-UML}[PRIMAmob-UML]{mobile \acs{PRIMA-UML}}
\acrodef{PRIMA-UML}[PRIMA-UML]{PeRformance IncreMental vAlidation in \acs{UML}}
\acrodef{PROM}[PROM]{Programmable \acs{ROM}}
\acrodef{PROMISE}[PROMISE]{Participative Research labOratory  for Multimedia and Multilingual Information Systems Evaluation}
\acrodef{pSQL}[pSQL]{propagate \acs{SQL}}
\acrodef{PUI}[PUI]{Participant User Interface}
\acrodef{PURL}[PURL]{Persistent \acs{URL}}
\acrodef{QA}[QA]{Question Answering}
\acrodef{QE}[QE]{Query Expansion}
\acrodef{QoS-UML}[QoS-UML]{\acs{UML} Profile for QoS and Fault Tolerance}
\acrodef{QPA}[QPA]{Query Performance Analyzer}
\acrodef{QPP}[QPP]{Query Performance Prediction}
\acrodef{R}[R]{Recall}
\acrodef{RAM}[RAM]{Random Access Memory}
\acrodef{RAMM}[RAM]{Random Access Machine}
\acrodef{RBO}[RBO]{Rank-Biased Overlap}
\acrodef{RBP}[RBP]{Rank-Biased Precision}
\acrodef{RBTO}[RBTO]{Rank-Based Total Order}
\acrodef{RDBMS}[RDBMS]{Relational \acs{DBMS}}
\acrodef{RDF}[RDF]{Resource Description Framework}
\acrodef{REST}[REST]{REpresentational State Transfer}
\acrodef{REV}[REV]{Remote Evaluation}
\acrodef{RF}[RF]{Relevance Feedback}
\acrodef{RFC}[RFC]{Request for Comments}
\acrodef{RIA}[RIA]{Reliable Information Access}
\acrodef{RM}[RM]{Relevance Model}
\acrodef{RMSE}[RMSE]{Root Mean Square Error}
\acrodef{RMT}[RMT]{Run Management Tool}
\acrodef{ROM}[ROM]{Read Only Memory}
\acrodef{ROMIP}[ROMIP]{Russian Information Retrieval Evaluation Seminar}
\acrodef{RoMP}[RoMP]{Rankings of Measure Pairs}
\acrodef{RoS}[RoS]{Rankings of Systems}
\acrodef{RP}[RP]{Relative Position}
\acrodef{RR}[RR]{Reciprocal Rank}
\acrodef{RSM}[RSM]{Run Storing Manager}
\acrodef{RSS}[RSS]{Sum of Squares of Residuals}
\acrodef{RST}[RST]{Rhetorical Structure Theory}
\acrodef{RSV}[RSV]{Retrieval Status Value}
\acrodef{RT-UML}[RT-UML]{\acs{UML} Profile for Schedulability, Performance and Time}
\acrodef{SA}[SA]{Software Architecture}
\acrodef{SAL}[SAL]{Storing Abstraction Layer}
\acrodef{SAMT}[SAMT]{Statistical Analysis Management Tool}
\acrodef{SAN}[SAN]{Sistema Archivistico Nazionale}
\acrodef{SASM}[SASM]{Statistical Analysis Storing Manager}
\acrodef{sARE}[sARE]{scaled Absolute Rank Error}
\acrodef{sARE(AP)}[sARE-AP]{AP induced scaled Absolute Rank Error}
\acrodef{SBTO}[SBTO]{Set-Based Total Order}
\acrodef{SCS}[SCS]{Simplified query Clarity Score}
\acrodef{SCQ}[SCQ]{Similarity Collection-Query}
\acrodef{SD}[SD]{Sequence Diagram}
\acrodef{SE}[SE]{Search Engine}
\acrodef{SEBD}[SEBD]{Convegno Nazionale su Sistemi Evoluti per Basi di Dati}
\acrodef{SEM}[SEM]{Standard Error of the Mean}
\acrodef{SERP}[SERP]{Search Engine Result Page}
\acrodef{SFT}[SFT]{Satisfaction--Frustration--Total}
\acrodef{SIL}[SIL]{Service Integration Layer}
\acrodef{SIP}[SIP]{Submission Information Package}
\acrodef{SKOS}[SKOS]{Simple Knowledge Organization System}
\acrodef{SM}[SM]{Software Model}
\acrodef{sMARE}[sMARE]{scaled Mean Absolute Rank Error}
\acrodef{SME}[SME]{Statistics--Metrics-Experiments}
\acrodef{SMART}[SMART]{System for the Mechanical Analysis and Retrieval of Text}
\acrodef{SMV}[SMV]{Score Magnitude and Variance}
\acrodef{SoA}[SoA]{Service-oriented Architectures}
\acrodef{SOA}[SOA]{Strength of Association}
\acrodef{SOAP}[SOAP]{Simple Object Access Protocol}
\acrodef{SOM}[SOM]{Self-Organizing Map}
\acrodef{SPARQL}[SPARQL]{Simple Protocol and RDF Query Language}
\acrodef{SPE}[SPE]{Software Performance Engineering}
\acrodef{SPINA}[SPINA]{Superimposed Peer Infrastructure for iNformation Access}
\acrodef{SPLIT}[SPLIT]{Stemming Program for Language Independent Tasks}
\acrodef{SPOOL}[SPOOL]{Simultaneous Peripheral Operations On Line}
\acrodef{SQL}[SQL]{Structured Query Language}
\acrodef{SR}[SR]{Sliding Ratio}
\acrodef{sRBP}[sRBP]{Session Rank Biased Precision}
\acrodef{SRU}[SRU]{Search/Retrieve via \acs{URL}}
\acrodef{SS}[SS]{Sum of Squares}
\acrodef{SSTF}[SSTF]{Shortest Seek Time First}
\acrodef{STAR}[STAR]{Steiner-Tree Approximation in Relationship graphs}
\acrodef{STON}[STON]{STemming ON}
\acrodef{SVM}[SVM]{Support Vector Machine}
\acrodef{TAC}[TAC]{Text Analysis Conference}
\acrodef{TBG}[TBG]{Time-Biased Gain}
\acrodef{TCP}[TCP]{Transmission Control Protocol}
\acrodef{TEL}[TEL]{The European Library}
\acrodef{TERRIER}[TERRIER]{TERabyte RetrIEveR}
\acrodef{TF}[TF]{Term Frequency}
\acrodef{TFR}[TFR]{True False Rate}
\acrodef{TIR}[TIR]{Traditional IR}
\acrodef{TLD}[TLD]{Top Level Domain}
\acrodef{TME}[TME]{Topics--Metrics-Experiments}
\acrodef{TN}[TN]{True Negative}
\acrodef{TO}[TO]{Transfer Object}
\acrodef{TP}[TP]{True Positve}
\acrodef{TPR}[TPR]{True Positive Rate}
\acrodef{TRAT}[TRAT]{Text Relevance Assessing Task}
\acrodef{TREC}[TREC]{Text REtrieval Conference}
\acrodef{TRECVID}[TRECVID]{TREC Video Retrieval Evaluation}
\acrodef{TTL}[TTL]{Time-To-Live}
\acrodef{UCD}[UCD]{Use Case Diagram}
\acrodef{UDC}[UDC]{Universal Decimal Classification}
\acrodef{UEF}[UEF]{Utility Estimation Framework}
\acrodef{uGAP}[uGAP]{User-oriented Graded Average Precision}
\acrodef{UI}[UI]{User Interface}
\acrodef{UML}[UML]{Unified Modeling Language}
\acrodef{UMT}[UMT]{User Management Tool}
\acrodef{UMTS}[UMTS]{Universal Mobile Telecommunication System}
\acrodef{UoM}[UoM]{Utility-oriented Measurement}
\acrodef{UPC}[UPC]{Universal Product Code}
\acrodef{URI}[URI]{Uniform Resource Identifier}
\acrodef{URL}[URL]{Uniform Resource Locator}
\acrodef{URN}[URN]{Uniform Resource Name}
\acrodef{USM}[USM]{User Storing Manager}
\acrodef{VA}[VA]{Visual Analytics}
\acrodef{VAIRE}[VAIR\"{E}]{Visual Analytics for Information Retrieval Evaluation}
\acrodef{VATE}[VATE$^2$]{Visual Analytics Tool for Experimental Evaluation}
\acrodef{VIRTUE}[VIRTUE]{Visual Information Retrieval Tool for Upfront Evaluation}
\acrodef{VD}[VD]{Virtual Document}
\acrodef{VDM}[VDM]{Visual Data Mining}
\acrodef{VIAF}[VIAF]{Virtual International Authority File}
\acrodef{VIM}[VIM]{International Vocabulary of Metrology}
\acrodef{VL}[VL]{Visual Language}
\acrodef{VoIP}[VoIP]{Voice over IP}
\acrodef{VS}[VS]{Visual Sentence}
\acrodef{W3C}[W3C]{World Wide Web Consortium}
\acrodef{WAN}[WAN]{Wide Area Network}
\acrodef{WIG}[WIG]{Weighted Information Gain}
\acrodef{WHO}[WHO]{World Health Organization}
\acrodef{WLAN}[WLAN]{Wireless \acs{LAN}}
\acrodef{WP}[WP]{Work Package}
\acrodef{WS}[WS]{Web Services}
\acrodef{WSD}[WSD]{Word Sense Disambiguation}
\acrodef{WSDL}[WSDL]{Web Services Description Language}
\acrodef{WWW}[WWW]{World Wide Web}
\acrodef{XMI}[XMI]{\acs{XML} Metadata Interchange}
\acrodef{XML}[XML]{eXtensible Markup Language}
\acrodef{XPath}[XPath]{XML Path Language}
\acrodef{XSL}[XSL]{eXtensible Stylesheet Language}
\acrodef{XSL-FO}[XSL-FO]{\acs{XSL} Formatting Objects}
\acrodef{XSLT}[XSLT]{\acs{XSL} Transformations}
\acrodef{YAGO}[YAGO]{Yet Another Great Ontology}
\acrodef{YASS}[YASS]{Yet Another Suffix Stripper}
\acrodef{AIC}[AIC]{Akaike Information Criterion}
\acrodef{BIC}[BIC]{Bayesian Information Criterion}
\acrodef{ssd}[ssd]{statistically significantly different}

\acrodef{AA}[AA]{Active Agreements}
\acrodef{MA}[MA]{Mixed Agreements}
\acrodef{PA}[PA]{Passive Agreements}
\acrodef{PAA}[PPA]{Proportion of Active Agreements}
\acrodef{PAD}[PPD]{Proportion of Active Disagreements}
\acrodef{PD}[PD]{Passive Disagreements}
\acrodef{PPA}[PPA]{Proportion of Passive Agreements}
\acrodef{PPD}[PPD]{Proportion of Passive Disagreements}
\acrodef{MD}[MD]{Mixed Disagreements}
\acrodef{AD}[AD]{Active Disagreements}

\newcommand{\robustIV}{Robust `04\xspace}
\newcommand{\dplrnpassXIX}{Deep Learning `19\xspace}

\begin{document}
\title{Query Performance Prediction for Neural IR:\\Are We There Yet?}

\titlerunning{QPP for Neural IR: Are We There Yet?}
\author{Guglielmo Faggioli\inst{1} \and
Thibault Formal\inst{2,3} \and
Stefano Marchesin \inst{1} \and
Stéphane Clinchant\inst{2} \and
Nicola Ferro \inst{1} \and
Benjamin Piwowarski\inst{3,4}
}

\authorrunning{G. Faggioli et al.}

\institute{University of Padova, Padova, Italy \and Naver Labs Europe, Meylan, France \and Sorbonne Université, ISIR Paris, France \and CNRS}

\maketitle              
\begin{abstract}
Evaluation in \ac{IR} relies on post-hoc empirical procedures, which are time-consuming and expensive operations. To alleviate this, \ac{QPP} models have been developed to estimate the performance of a system without the need for human-made relevance judgements. Such models, usually relying on lexical features from queries and corpora, have been applied to traditional sparse IR methods -- with various degrees of success. 
With the advent of neural IR and large Pre-trained Language Models, the retrieval paradigm has significantly shifted towards more semantic signals. In this work, we study and analyze to what extent current \ac{QPP} models can predict the performance of such systems. Our experiments consider seven traditional bag-of-words and seven BERT-based IR approaches, as well as nineteen state-of-the-art \acp{QPP} evaluated on two collections, Deep Learning '19 and Robust '04. 
Our findings show that \acp{QPP} perform statistically significantly worse on neural \ac{IR} systems. In settings where semantic signals are prominent (e.g., passage retrieval), their performance on neural models drops by as much as 10\% compared to bag-of-words approaches. On top of that, in lexical-oriented scenarios, \acp{QPP} fail to predict performance for neural \ac{IR} systems on those queries where they differ from traditional approaches the most.
\end{abstract}

\section{Introduction}
\label{sec:intro}


The advent of \acf{NIR} and \acf{PLM} induced considerable changes in several central \ac{IR} research and application areas, with implications that are yet to be fully tamed by the research community.
\acf{QPP} is defined as the prediction of the performance of an \ac{IR} system without human-crafted relevance judgements and is one of the areas the most interested by advancements in \ac{NIR} and \ac{PLM} domains. 
In fact, \emph{i)} \ac{PLM} can help developing better \ac{QPP} models, and \emph{ii)} it is not fully clear yet whether current \ac{QPP} techniques can be successfully applied to \ac{NIR}. 
With this paper, we aim to explore the connection between \ac{PLM}-based first-stage retrieval techniques and the available \ac{QPP} models. We are interested in investigating to what extent \ac{QPP} techniques can be applied to such \ac{IR} systems, given \begin{enumerate*}[label=\textit{\roman*})]
\item their fundamentally different underpinnings compared to traditional lexical \ac{IR} approaches,
\item that they hold the promise to replace -- or at least complement -- them in multi-stage ranking pipelines.
\end{enumerate*}
In return, \ac{QPP} advantages are multi-fold: it can be used to select the best-performing system for a given query, help users in reformulating their needs, or identify pathological queries that require manual intervention from the system administrators. Said otherwise, the need for QPP still holds for \ac{NIR} methods. 
Among the plethora of available \ac{QPP} methods, most of them rely on lexical aspects of the query and the collection. Such approaches have been devised, tested, and evaluated in predicting the performance of lexical bag-of-words \ac{IR} systems -- from now on referred to as \ac{TIR} -- with various degrees of success.
Recent advances in \ac{NLP} led to the advent of \ac{PLM}-based \ac{IR} systems, which shifted the retrieval paradigm from traditional approaches based on lexical matching to exploiting contextualized semantic signals -- thus alleviating the semantic gap problem. To ease the readability throughout the rest of the manuscript, with an abuse of notation, we use the more general term \ac{NIR} to explicitly refer to first-stage \ac{IR} systems based on BERT~\cite{devlin_etal-2018}. 

At the current time, 
no large-scale work has been devoted to assessing whether traditional \ac{QPP} models can be used for \ac{NIR} systems -- which is the goal of this study. 
We compare the performance of nineteen \ac{QPP} methods applied to seven traditional \ac{TIR} systems, with those achieved on seven state-of-the-art first-stage \ac{NIR} approaches based on \ac{PLM}. 
We consider both pre- and post-retrieval \ac{QPP}s, and include in our analyses post-retrieval \ac{QPP} models that exploit lexical or semantic signals to compute their predictions.
To instantiate our analyses on different scenarios we consider two widely adopted experimental collections: \robustIV and \dplrnpassXIX. 
Our contributions are as follows:
\begin{itemize}
    \item we apply and evaluate several state-of-the-art \ac{QPP} approaches to multiple \ac{NIR} retrievers based on BERT, on \robustIV and \dplrnpassXIX;
    \item we observe a correlation between \acp{QPP} performance and how different \ac{NIR} architectures perform lexical match;
    \item we show that currently available \acp{QPP} perform reasonably well when applied to \ac{TIR} systems, while they fail to properly predict the performance for \ac{NIR} systems, even on \ac{NIR} oriented collections;
    \item we highlight how such decrease in \ac{QPP} performance is particularly prominent on queries where \ac{TIR} and \ac{NIR} performances differ the most -- which are those queries where \acp{QPP} would be most beneficial.
\end{itemize}

The remainder of this paper is organized as follows: Section~\ref{sec:related} outlines the main related endeavours. Section~\ref{sec:methods} details our methodology, while Section~\ref{sec:setup} contains the experimental setting. Empirical results are reported in Section~\ref{sec:results}. Section~\ref{sec:concs} summarizes the main conclusions and future research directions.

\section{Related Work}
\label{sec:related}

The rise of large \ac{PLM} like BERT~\cite{devlin_etal-2018} has given birth to a new generation of \ac{NIR} systems. Initially employed as re-rankers in a standard learning-to-rank framework~\cite{NogueiraCho2019}, a real paradigm shift occurred when the first \ac{PLM}-based retrievers outperformed standard \ac{TIR} models as candidate generators in a multi-stage ranking setting. 
For such a task, dense representations, based on a simple pooling of contextualized embeddings, 
combined with approximate nearest neighbors algorithms, have proven to be both highly effective and efficient~\cite{ReimersGurevych2019,karpukhin-etal-2020-dense,xiong2021approximate,Hofstaetter2021_tasb_dense_retrieval,gao-callan-2022-unsupervised,izacard2021contriever}. 
ColBERT~\cite{khattab_etal-2020,SanthanamEtAl2021} avoids this pooling mechanism, and directly models semantic matching at the token level -- allowing it to capture finer-grained relevance signals. In the meantime, another research branch brought lexical models up to date, by taking advantage of BERT and the proven efficiency of inverted indices in various manners. Such sparse approaches for instance learn contextualized term weights~\cite{DaiCallan2020,MalliaEtAl2021,ZhuangZuccon2021,LinMa2021}, query or document expansion~\cite{NogueiraEtAl2019}, or both mechanisms jointly~\cite{FormalEtAl2021,FormalEtAl2022}. This new wave of \ac{NIR} systems, which substantially differ from lexical ones -- and from each other -- demonstrate state-of-the-art results on several datasets, from MS MARCO~\cite{BajajEtAl2016} on which models are usually trained, to zero-shot settings such as the BEIR~\cite{thakur2021beir} or LoTTE~\cite{SanthanamEtAl2021} benchmarks. 





A well-known problem linked to \ac{IR} evaluation is the variation in performance achieved by different \ac{IR} systems, even on a single query~\cite{CarmelYomTov2010,CulpepperEtAl2021b}. To partially account for it, a large body of work has focused on predicting the performance that a system would achieve for a given query, using \ac{QPP} models. Such models are typically divided into pre- and post-retrieval predictors.
Traditional pre-retrieval \acp{QPP} leverage statistics on the query terms occurrences~\cite{HauffEtAl2008}. 
For example, SCQ~\cite{Zhao2008EffectiveEvidence}, VAR~\cite{Zhao2008EffectiveEvidence} and IDF~\cite{Cronen-Townsend2004AExpansion,Scholer2004QuerySearch} combine query tokens' occurrence indicators, such as \ac{CF} and \ac{IDF}, to compute their performance prediction score.
Post-retrieval \acp{QPP} exploit the results of \ac{IR} models for the given query~\cite{CarmelYomTov2010}. Among them, Clarity~\cite{Cronen-Townsend2002PredictingPerformance} 
compares the language model of the first $k$ retrieved documents with the one of the entire corpus.
NQC~\cite{Shtok2012PredictingEstimation}, WIG~\cite{Zhou2007QueryEnvironments} and SMV~\cite{Tao2014QueryTogether} exploit the retrieval scores distribution for the top-ranked documents to compute their predictive score.
Finally, \ac{UEF}~\cite{ShtokEtAl2010} serves as a general framework that can be instantiated with many of the mentioned predictors, pre-retrieval ones included. Post-retrieval predictors are based on lexical signals -- SMV, NQC and WIG rely on the Language Model scores estimated from top-retrieved documents, while Clarity and UEF exploit the language models of the top-$k$ documents.

We further divide \ac{QPP} models into traditional and neural approaches. 
Among neural predictors, one of the first approaches is NeuralQPP~\cite{zamani_etal-2018} which computes its predictions by combining semantic and lexical signals using a feed-forward neural network.
Notice that NeuralQPP is explicitly designed for \ac{TIR} and is hence not expected to work better with \ac{NIR}~\cite{zamani_etal-2018}.
A similar approach for Question Answering is NQA-QPP~\cite{hashemi_etal-2019}, which also relies on three neural components but, unlike NeuralQPP, exploits BERT~\cite{devlin_etal-2018} to embed tokens semantics. Similarly, BERT-QPP~\cite{arabzadeh_etal-2020c} encodes semantics via BERT, but directly \emph{fine-tunes} it to predict query performance based on the first retrieved document.
Subsequent approaches extend BERT-QPP by employing a groupwise predictor to jointly learn from multiple queries and documents~\cite{chen_etal-2022} or by transforming its pointwise regression into a classification task~\cite{datta_etal-2022c}. Since we did not consider multiple formulations, we did not experiment with such approach in our empirical evaluation.

Although traditional \ac{QPP} methods have been widely used over the years, only few works have been done to apply them on \ac{NIR} models. Similarly, neural \ac{QPP} methods -- which model the semantic interactions between query and document terms -- have been mostly \emph{designed for} and \emph{evaluated on} \ac{TIR} models.
Two noteworthy exceptions concerning the tested IR models are \cite{hashemi_etal-2019} who evaluate the devised \ac{QPP} on pre-BERT approaches for \ac{QA}, while~\cite{datta_etal-2022a} assess the performance of their approach on DRMM~\cite{guo_etal-2016} (pre-BERT) and ColBERT~\cite{khattab_etal-2020} (BERT-based) as \ac{NIR} models. 
Hence, there is an urgent need to deepen the evaluation of \ac{QPP} on state-of-the-art \ac{NIR} models to understand where we are, what are the challenges, and which directions are more promising.

A third category that can be considered a hybrid between the groups of predictors mentioned above is passage retrieval QPP~\cite{Roitman2018}. In \cite{Roitman2018}, authors exploit lexical signals obtained from passages' language models to devise a predictor meant to better deal with passage retrieval prediction.

\section{Methodology}
\label{sec:methods}

\paragraph{\textbf{Evaluating Query Performance Predictors.}}
\ac{QPP} models compute a score for each query, that is expected to correlate with the quality of the retrieval for such query.
Traditional evaluation of \ac{QPP} models relies on measuring the correlation between the predicted QPP scores and the observed performance measured with a traditional \ac{IR} measure. Typical correlation coefficients include Kendall's $\tau$, Spearman's $\rho$ and the Pearson's $r$.
This evaluation procedure has the drawback of summarizing, through the correlation score, the performance of a \ac{QPP} model into a single observation for each system and  collection~\cite{faggioli_etal_2021,FaggioliEtAl2022b}. Therefore, Faggioli et al.~\cite{faggioli_etal_2021} propose a novel evaluation approach based on the \ac{sARE} measure that, given a query $q$, is defined as $\operatorname{sARE}(q) = \frac{|R^e_q - R^p_q|}{|Q|}$, where $R^e_q$ and $R^p_q$ are the ranks of the query $q$ induced by the \ac{IR} measure and the \ac{QPP} score respectively, over the entire set of queries of size $|Q|$. With ``rank'' we refer to the ordinal position of the query if we sort all the queries of the collection either by \ac{IR} performance or prediction score.
By switching from a single-point estimation to a distribution of performance, \ac{sARE} has the advantage of allowing conducting more powerful statistical analyses and carrying out failure analyses on queries where the predictors are particularly bad. 
To be comparable with previous literature, we report in Sec.~\ref{subsec:qpp_models_perfrormance} the performance of the analyzed predictors using the traditional Pearson's $r$ correlation-based evaluation. On the other hand, we use \ac{sARE} as the evaluation measure for the statistical analyses, to exploit its additional advantages. Such analyses, whose results are reported in Sec.~\ref{subsec:anova_analysis}, are described in the remainder of this section.

\paragraph{\textbf{ANOVA.}}
To assess the effect induced by \ac{NIR} systems on \ac{QPP} performance, we employ the following \ac{ANOVA} models.
The first model, dubbed~\ref{eq:MD1}, aims at explaining the sARE performance given the predictor, the type of \ac{IR} model and the collection. Therefore, we define it as follows:

\begin{equation}
    \label{eq:MD1}\tag{MD1}
    sARE_{ijpqr} = \mu + \pi_p + \eta_i + \chi_j + (\eta\chi)_{ij} + \epsilon_{ijpqr},
\end{equation}
where $\mu$ is the grand mean, $\pi_p$ is the effect of the $p$-th predictor, $\eta_i$ represents the type of \ac{IR} model (either \ac{TIR} or \ac{NIR}), $\chi_j$ stands for the effect of the $j$-th collection on \ac{QPP}'s performance, and $(\eta\chi)_{ij}$ describes how much the type of run and the collection interact and $\epsilon$ is the associated error.

Secondly, since we are interested in determining the effect of different predictors in interaction with each query, we define a second model, dubbed~\ref{eq:MD2}, that also includes the interaction factor and is formulated as follows:
\begin{equation}
    \label{eq:MD2}\tag{MD2}
    sARE_{ipqr} = \mu + \pi_p + \tau_q + \eta_i + (\pi\tau)_{qp} + (\pi\eta)_{pi}+ (\tau\eta)_{iq} + \epsilon_{ipqr},
\end{equation}
Differently from \ref{eq:MD1}, we apply \ref{eq:MD2} to each collection separately. Therefore, having a single collection, we replace the effect of the collection with $\tau_q$, the effect for the $q$-th topic. Furthermore, the model includes also all the first-order interactions.

The \ac{SOA}~\cite{rutherford2011anova} is assessed using $\omega^2$ measure computed as: 
$$\omega^2_{<fact>} = \frac{\operatorname{df}_{<fact>}*F_{<fact>}}{\operatorname{df}_{<fact>}*(F_{<fact>}-1)*N},$$ 
where $N$ is the number of experimental data-points, $\operatorname{df}_{<fact>}$ is the factor's number of \ac{DF}, and $F_{<fact>}$ it the F statistics computed by \ac{ANOVA}. As a rule-of-thumb, $\omega^2<6\%$ indicates a small \ac{SOA}, $6\%\leq \omega^2<14\%$ is a medium-sized effect, while $\omega^2\geq 14\%$ represent a large-sized effect. 

ANOVA Models have been fitted using \texttt{anovan} function from the stats MATLAB package. In terms of sample size, depending on the model and collection at hand, we considered 19 predictors, 249 topics in the case of \robustIV and 43 for \dplrnpassXIX and 14 different \ac{IR} systems for a total of 66234 and 11438 observations for \robustIV and \dplrnpassXIX respectively.

\section{Experimental Setup}
\label{sec:setup}

Our analyses focus on two distinct collections: \robustIV~\cite{Voorhees2005}, and TREC Deep Learning 2019 Track (\dplrnpassXIX)~\cite{Craswell2021}. The collections have respectively 249 and 43 topics each and are based on TIPSTER and MS MARCO passages corpora. \robustIV is one of the most used collections to test lexical approaches, while providing a reliable benchmark for \ac{NIR} models~\cite{VoorheesEtAl2022} -- even though they struggle to perform well on this collection, especially when evaluated in a zero-shot setting~\cite{thakur2021beir}. \dplrnpassXIX concerns passage retrieval from natural questions -- the formulation of queries and the nature of the documents (passages) make the retrieval harder for \ac{TIR} approaches, while \ac{NIR} systems tend to have an edge in retrieving relevant documents. 

Our main objective is to assess whether existing \acp{QPP} are effective in predicting the performance of different state-of-the-art \ac{NIR} models. As reference points, we consider seven \ac{TIR} methods: Language Model with Dirichlet (\texttt{LMD}) and Jelinek–Mercer (\texttt{LMJM}) smoothing~\cite{Zhai2008}, BM25, vector space model~\cite{SaltonBuckley1988} (\texttt{TFIDF}), InExpB2~\cite{AmatiVanRijsbergen2002} (\texttt{InEB2}), Axiomatic F1-EXP~\cite{FangZhai2005} (\texttt{AxF1e}), and \ac{DFI}~\cite{KocabasEtAl2013}. \ac{TIR} runs have been computed using Lucene. 
For the \ac{NIR} methods, we focus on BERT-based first-stage models. 
We consider state-of-the-art models from the three main families of \ac{NIR} models, which exhibit different behavior, and thus might respond to \acp{QPP} differently. We consider \emph{dense} models, \begin{enumerate*}[label=\textit{\roman*})]
\item a ``standard'' bi-encoder (\texttt{bi}) trained with negative log-likelihood,
\item TAS-B~\cite{Hofstaetter2021_tasb_dense_retrieval} (\texttt{bi-tasb}) whose training relies on topic-sampling and knowledge distillation
\item and finally CoCondenser~\cite{gao-callan-2022-unsupervised} (\texttt{bi-cc}) and Contriever~\cite{izacard2021contriever} (\texttt{bi-ct}) which are based on contrastive pre-training
\end{enumerate*}. We also consider two models from the \emph{sparse} family: SPLADE~\cite{FormalEtAl2021} (\texttt{sp}) with default training strategy, and its improved version SPLADE++~\cite{FormalEtAl2021b,FormalEtAl2022} (\texttt{sp++}) based on distillation, hard-negative mining and pre-training. We finally consider the \emph{late-interaction} ColBERTv2~\cite{SanthanamEtAl2021} (\texttt{colb2}). Models are fine-tuned on the MS MARCO passage dataset; given the absence of training queries in \robustIV, they are evaluated in a zero-shot manner, similarly to previous settings~\cite{thakur2021beir,SanthanamEtAl2021}. 
Besides the bi-encoder we trained on our own, we rely on open-source weights available for every model. 
The advantage of considering multiple \ac{TIR} and \ac{NIR} models is that \textit{i)} we achieve more generalizable results: different models, either \ac{TIR} or \ac{NIR} perform the best in different scenarios and therefore our conclusions should be as generalizable as possible; \textit{ii)} it allows to achieve more statistical power in the experimental evaluation.
We focus our analyses on \ac{nDCG} with cutoff 10, as it is employed across NIR benchmarks consistently. This is not the typical setting for evaluating traditional \ac{QPP} -- which usually considers \ac{AP}@1000.  Nevertheless, given our objective -- determining how \ac{QPP} performs on settings where \ac{NIR} models can be used successfully -- we are also interested in selecting the most appropriate measure.

Concerning \ac{QPP} models, we select the most popular state-of-the-art approaches. In details, we consider 9 pre-retrieval models: \ac{SCS}~\cite{HeEtAl2008}, \ac{SCQ}~\cite{Zhao2008EffectiveEvidence}, VAR~\cite{Zhao2008EffectiveEvidence}, \ac{IDF} and \ac{ICTF}~\cite{Cronen-Townsend2004AExpansion,Scholer2004QuerySearch}. For \ac{SCS}, we use the sum aggregation, while for others we use max and mean, which empirically produce the best results. In terms of post-retrieval \ac{QPP} models, our experiments are based on Clarity~\cite{Cronen-Townsend2002PredictingPerformance}, \ac{NQC}~\cite{Shtok2012PredictingEstimation}, \ac{SMV}~\cite{Tao2014QueryTogether}, \ac{WIG}~\cite{Zhou2007QueryEnvironments} and their \ac{UEF}~\cite{ShtokEtAl2010} counterparts. Among post-retrieval predictors, we also include a supervised approach, BERT-QPP~\cite{arabzadeh_etal-2020c}, using both bi-encoder (\textit{bi}) and cross-encoder (\textit{ce}) formulations. 
We train BERT-QPP\footnote{We use the implementation provided at https://github.com/Narabzad/BERTQPP} for each \ac{IR} system on the MS MARCO training set, as proposed in~\cite{arabzadeh_etal-2020c}.
Similarly to what is done for \ac{NIR} models, we apply BERT-QPP models on \robustIV queries in a zero-shot manner.

\section{Experimental Results}
\label{sec:results}

\subsection{QPP models performance}
\label{subsec:qpp_models_perfrormance}

\begin{table}[tb]
\centering
\caption{nDCG@10 for the selected \ac{TIR} and \ac{NIR} systems. \ac{NIR} outperform traditional approaches on \dplrnpassXIX, and have comparable performance on \robustIV. }
\resizebox{\textwidth}{!}{
\begin{tabular}{lcccccccccccccc}
\toprule
 & \texttt{axF1e} & \texttt{BM25} & \texttt{LMD} & \texttt{LMJM} & \texttt{TFIDF} & \texttt{DFI} &\texttt{InEB2} &\texttt{bi} & \texttt{bi-tasb} & \texttt{bi-cc} & \texttt{bi-ct} & \texttt{sp} & \texttt{sp++} & \texttt{colbv2} \\
\midrule
\dplrnpassXIX & 0.45 & 0.48 & 0.45 &0.48 & 0.37 & 0.47 & 0.49 &  0.64 & 0.72 & 0.72 & 0.67 & 0.71 & 0.73 & 0.75 \\
\robustIV & 0.39 & 0.44 & 0.43 & 0.40 & 0.31 & 0.44 & 0.44 & 0.23 & 0.45 & 0.30 &  0.46 & 0.39 & 0.45 & 0.47 \\
\bottomrule
\end{tabular}}
\label{table:perf}
\end{table}

\begin{figure}[!tb]
     \centering
     \begin{subfigure}[b]{0.65\textwidth}
         \centering
         \includegraphics[width=\textwidth,trim={36em 5em 40em 9em},clip]{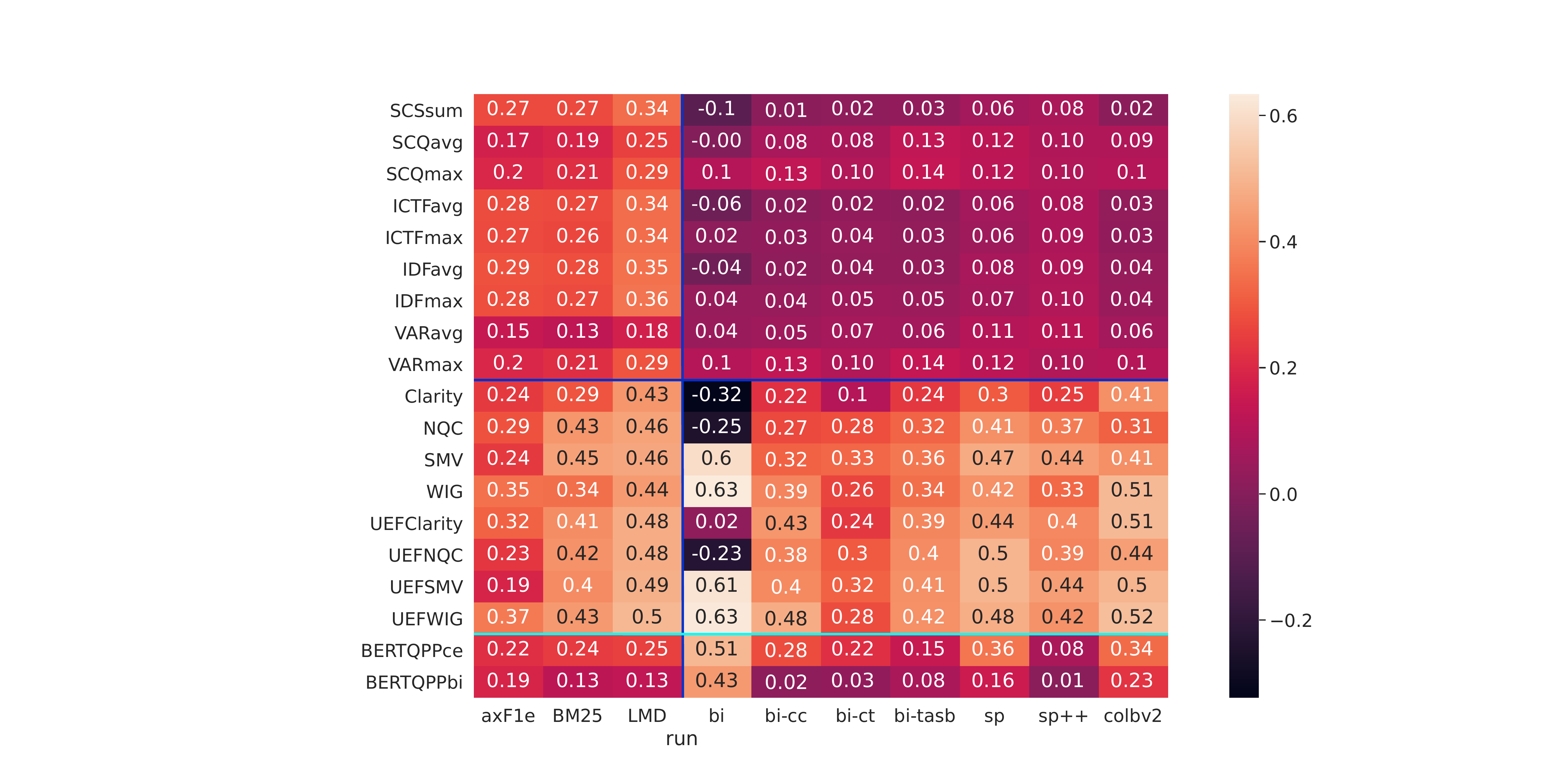}
         \caption{\robustIV}
         \label{fig:qpp_heatmap_robust04}
     \end{subfigure}
     \begin{subfigure}[b]{0.65\textwidth}
         \centering
         \includegraphics[width=\textwidth,trim={36em 5em 40em 9em},clip]{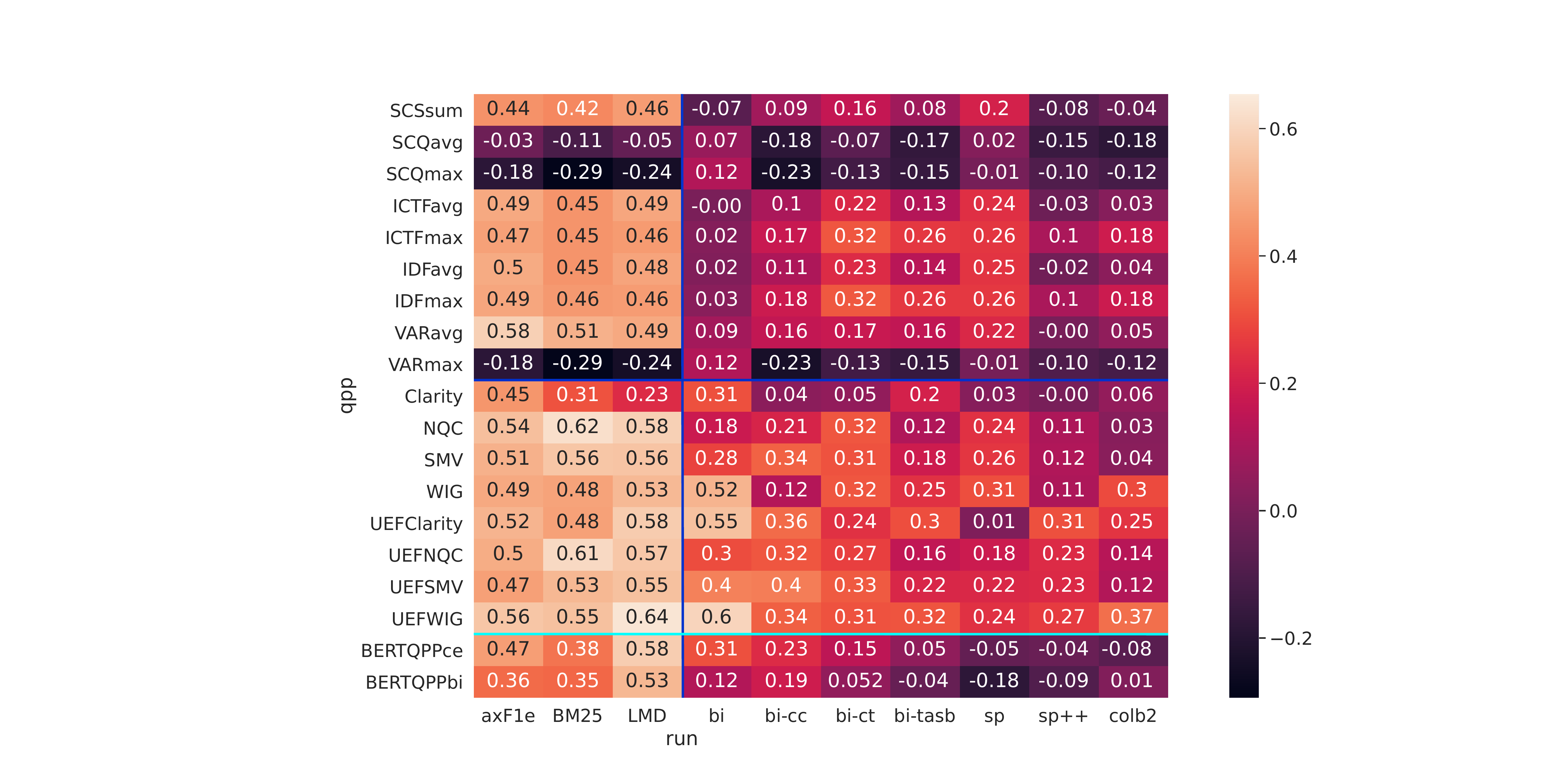}
         \caption{\dplrnpassXIX}
         \label{fig:qpp_heatmap_dplrnpass}
     \end{subfigure}
    \caption{Pearson's $r$ correlation observed for different pre (top) and post (bottom) retrieval predictors on lexical (left) and neural (right) runs. To avoid cluttering, we report the results for the 3 main \ac{TIR} models, other models achieve highly similar results.}
    \label{fig:qpp_heatmaps}
\vspace{-2em}
\end{figure}

Figures~\ref{fig:qpp_heatmap_robust04} and \ref{fig:qpp_heatmap_dplrnpass} refer, respectively, to \robustIV and \dplrnpassXIX collections and report the Pearson's $r$ correlation between the scores predicted by the chosen predictors and the \ac{nDCG}@10, for both \ac{TIR} and \ac{NIR} runs\footnote{Additional IR measures and correlations, as well as full ANOVA tables are available at: \url{https://github.com/guglielmof/ECIR2023-QPP}}. 
The presence of negative values indicates that some predictors fail in specific contexts and has been observed before in the \ac{QPP} setting~\cite{Hauff2010}.
For \robustIV, we notice that -- following previous literature -- pre-retrieval (top) predictors (mean correlation: 15.9\%) tend to perform 52.3\% worse than post-retrieval ones (bottom) (mean correlation: 30.2\%). Pre-retrieval results are in line with previous literature~\cite{ZendelEtAl2019}. The phenomenon is more evident (darker colors) for \ac{NIR} runs (right) than \ac{TIR} ones (left). 
Pre-retrieval predictors fail in predicting the performance of \ac{NIR} systems (mean correlation 6.2\% vs 25.6\% for \ac{TIR}), while in general, to our surprise, we notice that post-retrieval predictors tend to perform similarly on \ac{TIR} and \ac{NIR} (34.5\% vs 32.3\%) -- with some exceptions. For instance, for \texttt{bi}, post-retrieval predictors either perform extremely well or completely fail. This happens particularly on Clarity, NQC, and their UEF counterparts. Note that \texttt{bi} is the worst performing approach on \robustIV, with  23\% of nDCG@10 -- the second worst is \texttt{bi-cc} which achieves 30\% nDCG@10. 

The patterns observed for \robustIV hold only partially on \dplrnpassXIX. For example, we notice again that pre-retrieval predictors (mean correlation: 14.7\%) perform 58.3\% worse than post-retrieval ones (mean correlation: 35.3\%).
On the contrary, the difference in performance is far more evident between \ac{NIR} and \ac{TIR}. On \ac{TIR} runs, almost all predictors perform particularly well (mean correlation: 38.1\%) -- even better than on \robustIV collection. The only three exceptions are SCQ (both in avg and max formulations) and VAR using max formulation. Conversely, on \ac{NIR} the performance is overall lower (13.1\%) and relatively more uniform between pre- (5.4\%) and post-retrieval (19.9\%) models.
In absolute value, maximum correlation achieved by pre-retrieval predictors for \ac{NIR} on \dplrnpassXIX is much higher than the one achieved on \robustIV, especially for \texttt{bi-ct}, \texttt{sp}, and \texttt{bi-tasb} runs. On the other hand, post-retrieval predictors, perform worse than on the \robustIV. The only exception to this pattern is again represented by \texttt{bi}, on which some post-retrieval predictors, namely WIG, UEFWIG, and UEFClarity work surprisingly well. The supervised BERT-QPP shows a trend similar to other post-retrieval predictors on \dplrnpassXIX (42.3\% mean correlation against 52.9\% respectively) for what concerns \ac{TIR}, with performance in line with the one reported in~\cite{arabzadeh_etal-2020c}. This is exactly the setting where BERT-QPP has been devised and tested. If we focus on \dplrnpassXIX and \ac{NIR} systems, its performance (mean correlation: 4.5\%) is far lower than those of other post-retrieval predictors (mean correlation without BERT-QPP: 23.8\%). 
Finally, its performance on \robustIV~-- applied in zero-shot -- is considerably lower compared to other post-retrieval approaches.

Interestingly, on \robustIV, post-retrieval \acp{QPP} achieve, on average, top performance on the late interaction model (\texttt{colb2}), followed by sparse approaches (\texttt{sp} and \texttt{sp++}). Finally, excluding \texttt{bi}, where predictors achieve extremely inconsistent performance, dense approaches are those where \acs{QPP} perform the worst. 
In this sense, the performance that \ac{QPP} methods achieve on \ac{NIR} systems seems to correlate with the importance these systems give to lexical signals. In this regard, Formal et al.~\cite{FormalEtAl2022} 
observed how late-interaction and sparse architectures tend to rely more on lexical signals, compared to dense ones. 
\begin{table}[tb]
\centering
\caption{Pearson's $r$ \ac{QPP} performance for three versions of \texttt{sp++} applied on \robustIV, with varying degree of sparsity (\texttt{sp++$_{2}$} $\succ$ \texttt{sp++$_{1}$} $\succ$ \texttt{sp++$_{0}$} in terms of sparsity). The more ``lexical'' are the models, the better QPP performs. $\bold{d}_l$ and $\bold{q}_l$ represent respectively the average document/query sizes (i.e. non-zero dimensions in SPLADE) on \robustIV.}
\label{tbl:lexicality_vs_qpp}
\resizebox{0.7\textwidth}{!}{
\begin{tabular}{rccccccccc}
\toprule
\multicolumn{1}{c}{} &
 $\bold{d_l}$/$\bold{q_l}$ & Clarity                                             & NQC                                                 & SMV                                                 & WIG                                                 & UEFClarity                                          & UEFNQC                                             & UEFSMV                                              & UEFWIG                                             \\
\midrule
\texttt{sp++$_{2}$} & 55/22 & \cellcolor[HTML]{127622}{\color[HTML]{FFFFFF} 0.26} & \cellcolor[HTML]{DDE8CB}0.31                        & \cellcolor[HTML]{77BC65}0.46                        & \cellcolor[HTML]{127622}{\color[HTML]{FFFFFF} 0.42} & \cellcolor[HTML]{127622}{\color[HTML]{FFFFFF} 0.44} & \cellcolor[HTML]{127622}{\color[HTML]{FFFFFF} 0.4} & \cellcolor[HTML]{127622}{\color[HTML]{FFFFFF} 0.48} & \cellcolor[HTML]{127622}{\color[HTML]{FFFFFF} 0.5} \\
\texttt{sp++$_{1}$}  & 79/29 & \cellcolor[HTML]{DDE8CB}0.2                         & \cellcolor[HTML]{77BC65}0.34                        & \cellcolor[HTML]{127622}{\color[HTML]{FFFFFF} 0.47} & \cellcolor[HTML]{77BC65}0.35                        & \cellcolor[HTML]{DDE8CB}0.38                        & \cellcolor[HTML]{127622}{\color[HTML]{FFFFFF} 0.4}                        & \cellcolor[HTML]{77BC65}0.46                        & \cellcolor[HTML]{77BC65}0.43                       \\
\texttt{sp++$_{0}$}  & 204/45 & \cellcolor[HTML]{77BC65}0.25                        & \cellcolor[HTML]{127622}{\color[HTML]{FFFFFF} 0.37} & \cellcolor[HTML]{DDE8CB}0.44                        & \cellcolor[HTML]{DDE8CB}0.33                        & \cellcolor[HTML]{77BC65}0.4                         & \cellcolor[HTML]{DDE8CB}0.39                       & \cellcolor[HTML]{DDE8CB}0.44                        & \cellcolor[HTML]{DDE8CB}0.42\\
\bottomrule
\end{tabular}
}
\end{table}
To further corroborate this observation, we apply the predictors to three versions of SPLADE++ with various levels of sparsit as controlled by the regularization hyperparameter. Increasing the sparsity of representations leads to models that cannot rely as much on expansion -- emphasizing the importance given to lexical signals in defining the document ranking. Therefore, as a first approximation, we can deem sparser methods to be also more lexical. 
Given the low performance achieved by pre-retrieval \acp{QPP}, we focus this analysis on post-retrieval methods only. 
Table~\ref{tbl:lexicality_vs_qpp} shows the Pearson's $r$ for the considered predictors and different SPLADE++ versions. Interestingly, in the majority of the cases, \acp{QPP} perform the best for the sparser version (\texttt{sp++$_2$}), followed by \texttt{sp++$_1$} and \texttt{sp++$_0$} -- which is the one used in Fig.~\ref{fig:qpp_heatmaps}. There are a few switches, often associated with very close correlation values (SMV and UEFClarity).
Only one predictor, NQC, completely reverses the order.
This goes in favour of our hypothesis that indeed \ac{QPP} performance tends to correlate with the degree of lexicality of the \ac{NIR} approaches. Although not directly comparable, following this line of thought, \texttt{sp}, being handled better by \acp{QPP} (cfr. Fig.~\ref{fig:qpp_heatmap_robust04}), is more lexical than all the \texttt{sp++} versions considered: this is reasonable, given the different training methodology. Finally, \texttt{colb2}, being the method where \acp{QPP} achieve the best performance, might be the one that, at least for what concerns the \robustIV collection, gives the highest importance to lexical signals -- in line with what was observed in~\cite{FormalEtAl2021}.


\subsection{ANOVA Analysis}
\label{subsec:anova_analysis}

To further statistically quantify the phenomena observed in the previous subsection, we apply \ref{eq:MD1} to our data, considering both collections at once. From a quantitative standpoint, we notice that all the factors included in the model are statistically significant ($\operatorname{p-value}<10^{-4}$). 
In terms of \ac{SOA}, the collection factor has a small effect (0.02\%). The run type, on the other hand, impacts for $\omega^2=0.48\%$. Finally, the interaction between the collection and run type, although statistically significant, has a small impact on the performance ($\omega^2=0.05\%$): in both collections \acp{QPP} perform better on \ac{TIR} models.
All factors are significant but have small-size effects. This is in contrast with what was observed for the performance of \ac{IR} systems~\cite{FerroSilvello2018,CulpepperEtAl2021b}, where most of the \ac{SOA} range between medium to large. Nevertheless, it is in line with what was observed by Faggioli et al.~\cite{faggioli_etal_2021} for the performance \ac{QPP} methods, who showed that all the factors besides the topic are small to medium. A second observation is that it is likely that the small \acp{SOA} are due to a model unable to accrue for all the aspects of the problem -- more factors should be considered. Model~\ref{eq:MD2}, introducing also the topic effect, allows for further investigation of this hypothesis.

\begin{figure}[tb]
    \centering
    \includegraphics[width=0.45\columnwidth,trim={10em 25em 10em 28em}, clip]{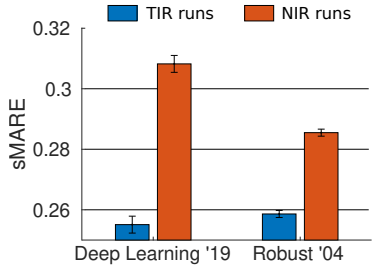}
    \caption{Comparison between the mean sARE (sMARE) achieved over \ac{TIR} or \ac{NIR} when changing the corpus. Observe the large distance between results on \ac{NIR} -- especially for \dplrnpassXIX~-- compared to the one on \ac{TIR} runs.}
    \label{fig:lex_vs_neu_performance}
\end{figure}

We are now interested in breaking down the performance of the predictors according to the collection and type of run.
Figure~\ref{fig:lex_vs_neu_performance} reports the average performance (measured with sMARE, the lower the better) for \acp{QPP} applied on \ac{NIR} or \ac{TIR} runs over different collections, with their confidence intervals as computed using ANOVA. 
Interestingly, regardless of the type of collection, the performance achieved by predictors on \ac{NIR} models will \emph{on average} be worse than those achieved on \ac{TIR} runs.
\ac{QPP} models perform better on \ac{TIR} than \ac{NIR} on both collections: this explains the small interaction effect between collections and run types. Secondly, there is no statistical difference between the performance achieved by \acp{QPP} applied to \ac{TIR} models when considering \dplrnpassXIX and \robustIV -- the confidence intervals are overlapping. This goes in contrast with what happens on \robustIV and \dplrnpassXIX when considering \ac{NIR} models: \acp{QPP} approaches applied on the latter dataset perform by far worse than on the former. 

\begin{table}[!tb]
\centering
\caption{p-values and $\omega^2$ \ac{SOA} using \ref{eq:MD2} on each collection}
\label{tbl:pvalues-MD3}

\begin{tabular}{r|cc|cc}
\toprule
\multicolumn{1}{l|}{} & \multicolumn{2}{c|}{\dplrnpassXIX} & \multicolumn{2}{c}{\robustIV}  \\
\multicolumn{1}{l|}{} & p-value  & $\omega^2$ & p-value & $\omega^2$ \\
\midrule
topic                & $<10^{-4}$ & 22.5\%     & $<10^{-4}$ & 24.0\%   \\
qpp                  & $<10^{-4}$ & 1.65\%     & $<10^{-4}$ & 2.21\%   \\
run type             & $<10^{-4}$ & 4.35\%     & $<10^{-4}$ & 0.11\%   \\
topic*qpp            & $<10^{-4}$ & 22.7\%     & $<10^{-4}$ & 17.2\%   \\
topic*run type       & $<10^{-4}$ & 15.2\%     & $<10^{-4}$ & 10.0\%   \\
qpp*run type         & 0.0012     & 0.23\%     & $<10^{-4}$ & 0.30\%   \\
\bottomrule
\end{tabular}
\end{table}

\begin{figure}[!tb]
     \centering
     \begin{subfigure}[b]{0.40\textwidth}
         \centering
         \includegraphics[width=\textwidth,trim={10em 25em 14em 28em},clip]{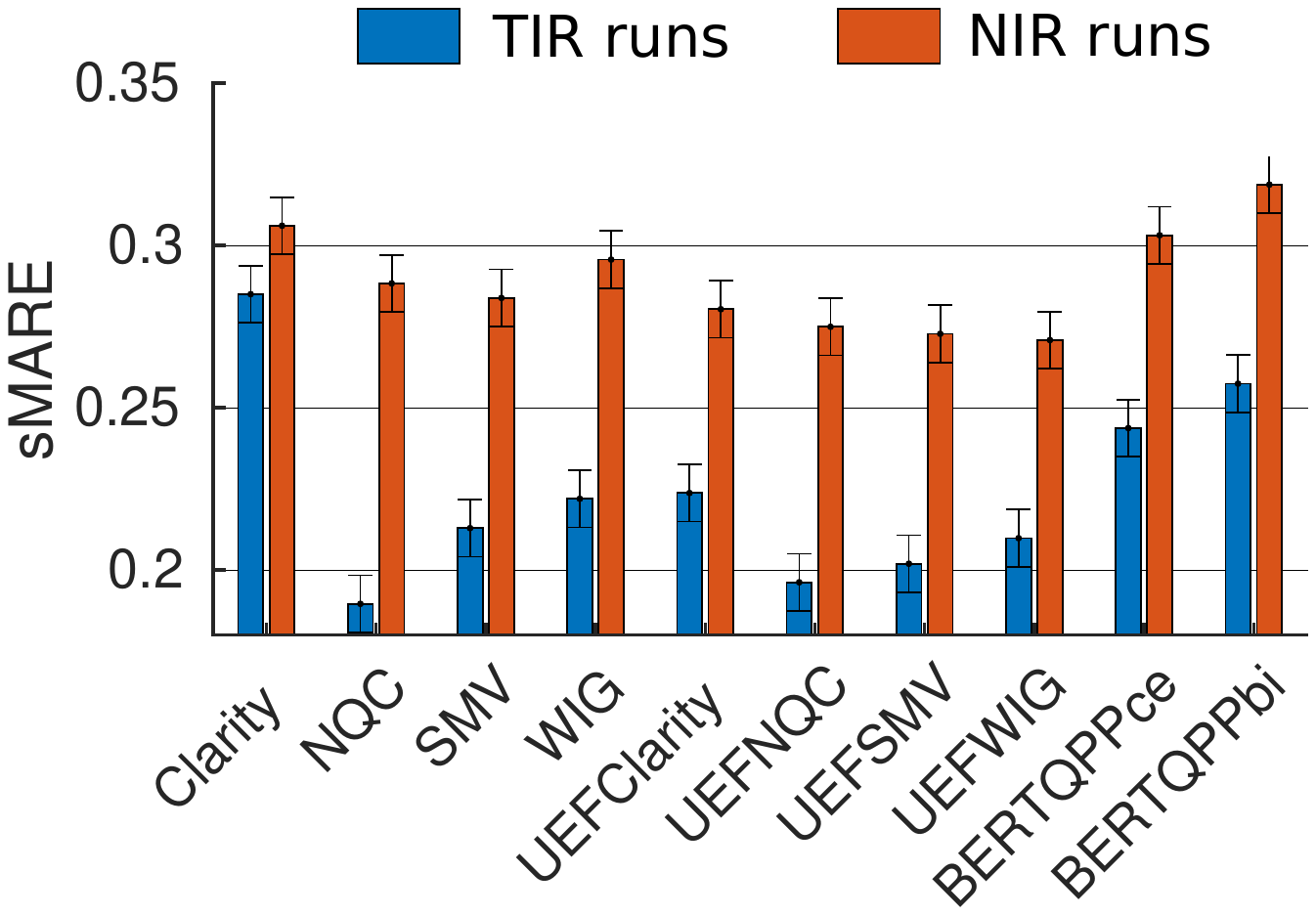}
         \caption{\dplrnpassXIX}
         \label{fig:qppwise-dplrnpass}
     \end{subfigure}
     \begin{subfigure}[b]{0.382\textwidth}
         \centering
         \includegraphics[width=\textwidth,trim={12em 25em 14em 28em},clip]{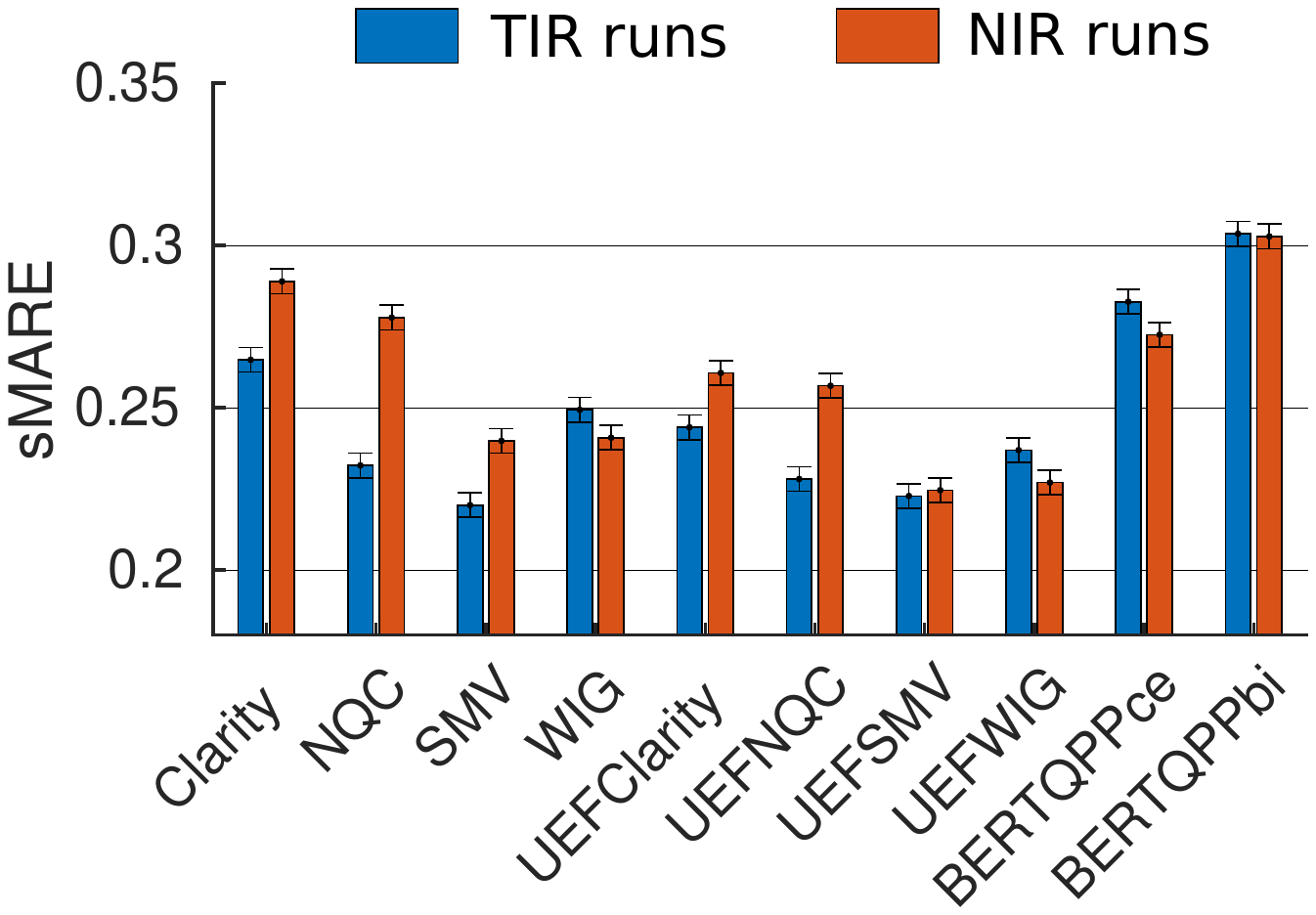}
         \caption{\robustIV}
         \label{fig:qppwise-RobustIV}
     \end{subfigure}
     \hfill
    \caption{sMARE observed for different predictors on \dplrnpassXIX (left) and \robustIV (right). On \dplrnpassXIX, predictors behave differently on \ac{TIR} and \ac{NIR} runs, while they are more uniform on \robustIV.}
    \label{fig:qppwise}
    \vspace{-2em}
\end{figure}

While \emph{on average} we will be less satisfied by \ac{QPP} predictors applied to \ac{NIR} regardless of the type of collection, there might be some noticeable exceptions of good performing predictors also for \ac{NIR} systems. To verify this hypothesis, we apply \ref{eq:MD2} to each collection separately, and measure what happens to each predictor individually\footnote{To avoid cluttering, we report the subsequent analyses only for post-retrieval predictors -- similar observations hold for pre-retrieval ones.}. Table~\ref{tbl:pvalues-MD3} reports the p-values and $\omega^2$ \ac{SOA} for the factors included in \ref{eq:MD2}, while Figure~\ref{fig:qppwise} depicts the phenomena visually. We observe that, concerning \dplrnpassXIX, the run type (\ac{TIR} or \ac{NIR}) is significant, while the interaction between the predictor and the run type is small: indeed predictors always perform better on \ac{TIR} runs than on \ac{NIR} ones. The only model that behaves slightly differently is Clarity, with far closer performance for both classes of runs -- this can be explained by the fact that Clarity is overall the worst-performing predictor. Notice that, the best predictor on \ac{TIR} runs -- NQC -- performs almost 10\% worse on \ac{NIR} ones.
Finally, we notice a large-size interaction between topics and \ac{QPP} models -- even bigger than the topic or \ac{QPP} themselves. This indicates that whether a model will be better than another strongly depends on the topic considered. An almost identical pattern was observed also in~\cite{faggioli_etal_2021}. Therefore, to improve \ac{QPP}'s generalizability, it is important not only to address challenges caused by differences in \ac{NIR} and \ac{TIR} but also to take into consideration the large variance introduced by topics. We analyze more in detail this variance later, where we consider only ``semantically defined'' queries.

If we consider \robustIV, the behaviour changes deeply: Figure~\ref{fig:qppwise} shows that predictors performances are much more similar for \ac{TIR} and \ac{NIR} runs compared to \dplrnpassXIX. This is further highlighted by the far smaller $\omega^2$ for run type on \robustIV in Table~\ref{tbl:pvalues-MD3} -- 4.35\% against 0.11\%. 
The widely different pattern between \dplrnpassXIX and \robustIV suggests that current \acp{QPP} are doomed to fail when used to predict the performance of \ac{IR} approaches that learned the semantics of a collection -- which is the case for \dplrnpassXIX that was used to fine-tune the models. 
Current \acp{QPP} evaluate better \ac{IR} approaches that rely on lexical clues. Such approaches include both \ac{TIR} models and \ac{NIR} models applied in a zero-shot fashion, as it is the case for \robustIV. 
Thus, \ac{QPP} models are expected to fail where \ac{NIR} models behave differently from the \ac{TIR} ones. 
This poses at stake one of the major opportunities provided by \ac{QPP}: if we fail in predicting the performance of \ac{NIR} models where they behave differently from \ac{TIR} ones, then a \ac{QPP} cannot be safely used to carry out model selection.
To further investigate this aspect, we carry out the following analysis: we select from \robustIV 25\% of the queries that are mostly ``semantically defined'' and rerun \ref{eq:MD2} on the new set of topics.
We call ``semantically defined'' those queries where \ac{NIR} behave, on average, oppositely w.r.t. the \ac{TIR}, either failing or succeeding at retrieving documents. In other terms, we select queries in the top quartile for the absolute difference in performance (nDCG), averaged over all \ac{TIR} or \ac{NIR} models. 


\begin{figure}[!tb]
     \centering
     \begin{subfigure}[b]{0.46\textwidth}
         \centering
         \includegraphics[width=\textwidth,trim={9.5em 25em 15em 28em},clip]{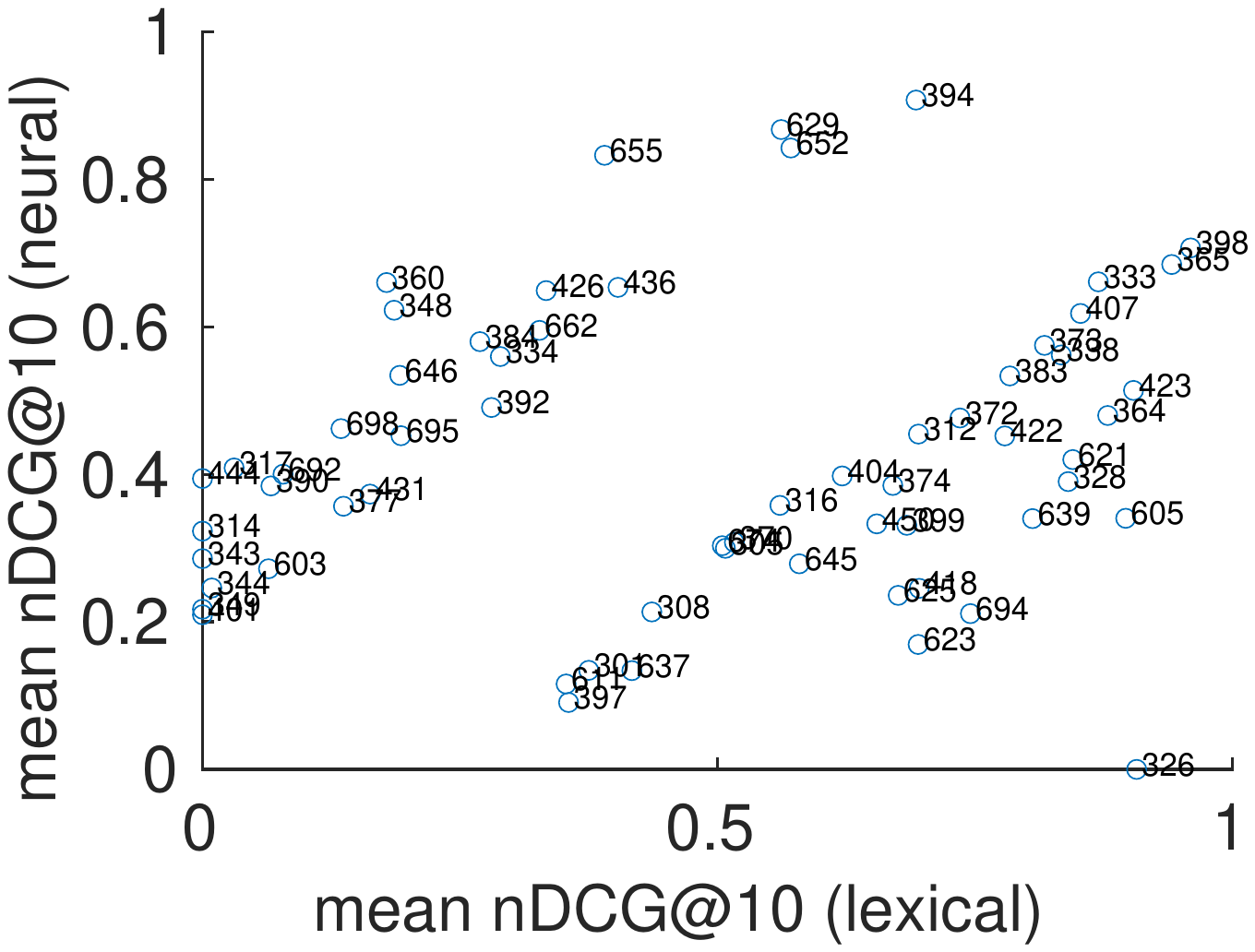}
         \caption{selected topics}
         \label{fig:selected-topics}
     \end{subfigure}
     \hfill
     \begin{subfigure}[b]{0.46\textwidth}
         \centering
         \includegraphics[width=\textwidth,trim={10em 25em 12em 28em},clip]{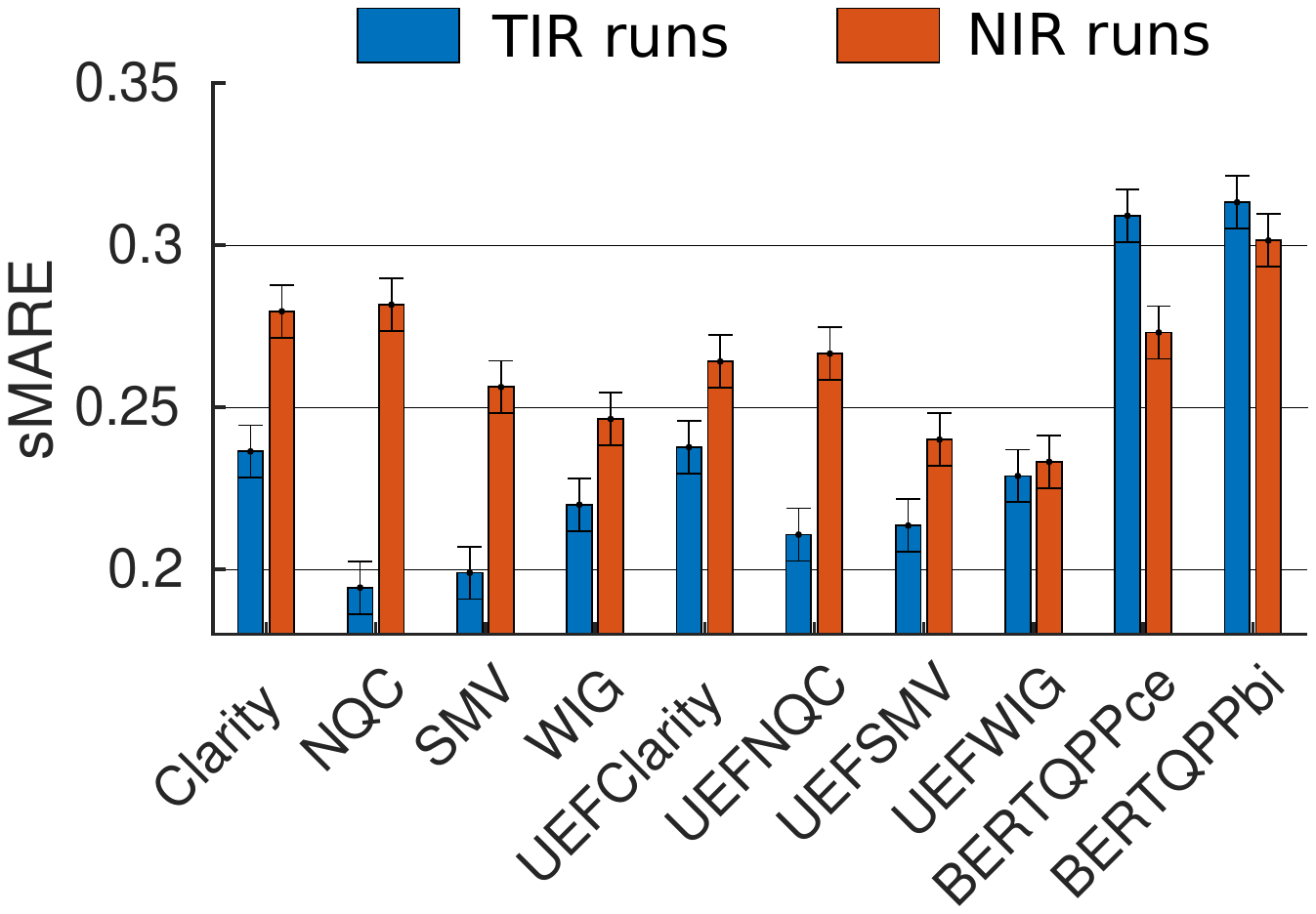}
         \caption{\robustIV}
         \label{fig:qppwise-RobustIV-tt}
     \end{subfigure}
    \caption{left: topics selected to maximize the difference between lexical and neural models; right: results of \ref{eq:MD2} applied on \robustIV considering only the selected topics.}
    \label{fig:qppwise-tt}
\end{figure}

Figure~\ref{fig:selected-topics} shows the performance of topics that maximize the difference between \ac{TIR} and \ac{NIR} and can be considered as more ``semantically defined''~\cite{FaggioliMarchesin21}. There are 62 topics selected (25\% of the 249 topics available on \robustIV). Of these, 35 topics are better handled by \ac{TIR} models, while 27 obtain better nDCG if dealt with \ac{NIR} rankers.
If we consider the results of applying \ref{eq:MD2} on this set of topics, we notice that compared to \robustIV (Table~\ref{tbl:pvalues-MD3}, last column) the effect of the different \acp{QPP} increases to 2.29\%: on these topics, there is more difference between different predictors.
The interaction between predictors and run types grows from 0.30\% to 0.91\%.
Furthermore, the effect of the run type grows from 0.11\% to 0.67\% -- 6 times bigger. On the selected topics, arguably those where a \ac{QPP} is the most useful to help select the right model, using \ac{NIR} systems has a negative impact (6 times bigger) on the performance of \acp{QPP}. 
Figure~\ref{fig:qppwise-RobustIV-tt}, compared to Figure \ref{fig:qppwise-RobustIV}, is more similar to Figure~\ref{fig:qppwise-dplrnpass} -- using only topics that are highly semantically defined, we get similar patterns as those observed for \dplrnpassXIX on Figure~\ref{fig:qppwise-dplrnpass}.
The only methods that behave differently are BERT-QPP approaches, whose performance is better on \ac{NIR} runs than on \ac{TIR} ones, but are the worst approaches in terms of predictive capabilities for both run types. In this sense, even though the contribution of the semantic signals appears to highly important to define new models with improved performance in the \ac{NIR} setting, it does not suffice to compensate for current \acp{QPP} limitations.

%
%
%

\section{Conclusion and Future Work}
\label{sec:concs}
With this work, we assessed to what extent current \acp{QPP}  are applicable to the recent family of first-stage \ac{NIR} models based on \ac{PLM}. To verify that, we evaluated 19 diverse \ac{QPP} models, used on seven traditional bag-of-words lexical models (\ac{TIR}) and seven first-stage \ac{NIR} methods based on BERT, applied to the \robustIV and \dplrnpassXIX collections.
We observed that if we consider a collection where \ac{NIR} systems had the chance to learn the semantics -- i.e., \dplrnpassXIX~-- \acp{QPP} are effective in predicting \ac{TIR} systems performance, but fail in dealing with \ac{NIR} ones. Secondly, we considered \robustIV. In this collection, \ac{NIR} models were applied in a zero-shot fashion, and thus behave similarly to \ac{TIR} models. In this case, we observed that \acp{QPP} tend to work better on \ac{NIR} models than in the previous scenario, but they fail on those topics where \ac{NIR} and \ac{TIR} models differ the most. This, in turn, impairs the possibility of using \ac{QPP} models to choose between \ac{NIR} and \ac{TIR} approaches where it is most needed.
On the other hand, semantic \ac{QPP} approaches such as BERT-QPP  do not solve the problem: being devised and tested on lexical IR systems, they work properly on such category of approaches but fail on neural systems. These results highlight the need for \acp{QPP} specifically tailored to Neural IR.

As future work, we plan to extend our analysis by considering other factors, such as the query variations to understand the impact that changing how a topic is formulated has on \ac{QPP}. Furthermore, we plan to devise \ac{QPP} methods explicitly designed to synergise with \ac{NIR} models, but that also take into consideration the large variance introduced by topics.

\subsubsection{Acknowledgements} 
The work was partially supported by University of Padova Strategic Research Infrastructure Grant 2017: “CAPRI: Calcolo ad Alte Pre-stazioni per la Ricerca e l’Innovazione”, ExaMode project, as part of the EU H2020 program under Grant Agreement no. 825292.

%
%
%
\newpage
\bibliographystyle{splncs04}
\bibliography{biblio}
\end{document}